\documentclass[aps,prl,twocolumn,groupedaddress]{revtex4}

\usepackage{latexsym}
\usepackage{amstext,amssymb,amsfonts}
\usepackage{graphicx}
\usepackage{color}
\usepackage{subfigure}

\begin{document}


\title{Explosive site percolation and finite size hysteresis}


\author{Nikolaos Bastas}
\email[]{nimpasta@physics.auth.gr}
\affiliation{Department of Physics, University of Thessaloniki, 54124 Thessaloniki, Greece}

\author{Kosmas Kosmidis}
\email[]{kkosm@physics.auth.gr}
\affiliation{Department of Physics, University of Thessaloniki, 54124 Thessaloniki, Greece}

\author{Panos Argyrakis}
\email[]{panos@physics.auth.gr}
\affiliation{Department of Physics, University of Thessaloniki, 54124 Thessaloniki, Greece}

\date{\today}

\begin{abstract}
We report the critical point for site percolation for the ``explosive'' type for 2D square lattices using Monte Carlo simulations and compare it to the classical well known percolation. We use similar algorithms as have been recently reported for bond percolation and networks. We calculate the ``explosive'' site percolation threshold as $p_c=0.695$ and we find evidence that  ``explosive'' site percolation surprisingly may belong to a different universality class than bond percolation on lattices, providing  that the transitions (a) are continuous and (b) obey  the conventional finite size scaling forms. We do not attempt to determine the order of the explosive transition.  Finally, we study and compare the direct and reverse processes, showing that while the reverse process is different from the direct process for finite size systems, the two cases become equivalent in the thermodynamic limit of large $L$.
\end{abstract}


\maketitle

\section{I.Introduction}
Random percolation (RP) \cite{Bunde_BOOK,Stauffer,Newman_Ziff1,Newman_Ziff2,Isichenko,HoshKop,Kirkpatrik} is a well studied model for phase transitions in statistical physics and condensed matter physics. It consists in randomly occupying sites (site percolation) or bonds (bond percolation) on a given lattice with probability $p$. Neighboring occupied sites merge to form clusters. As $p$ increases, new clusters are formed or coalesce, until at a critical value, $p_{c}$, a giant component emerges, transversing the system, the well-known infinite percolating cluster.\\
Classical percolation and all of its variants (continuum percolation, site-bond percolation, bootstrap percolation, invasion percolation, etc \cite{Bunde_BOOK}), are known to be continuous transitions, exhibiting a divergent correlation length $\xi$ at $p_{c}$. Thus, there was a stupendous surprise that recently, Achlioptas et al \cite{achlioptasDSouza2009}, proposed a new method for the occupation of sites which produces ``explosive'' transitions: when filling sequentially an empty lattice with occupied sites, instead of randomly occupying a site or bond, we choose two candidates and investigate which one of them leads to the smaller clustering. The one that does this is kept as a new occupied site on the lattice while the second one is discarded. This procedure considerably slows down the emergence of the giant component, which is now formed abruptly, thus the term ``explosive''.\\ 
Although it is not clear if ``explosive'' percolation is a first-order transition \cite{ziff2009explosive,Radicchi2010explosive,araujoHermann2010explosive} or continuous transition with a very small $\beta$ exponent \cite{Dorogovtsev2010}, it is certainly a very sharp transition with a lot of interesting and unusual properties and has stimulated considerable interest. Following these first announcements new work emerged \cite{Dorogovtsev2010,Grassberger2011} that doubted the result of the Achlioptas problem being a first-order phase transition, thus making this problem a hot contentious issue. Recently, applications of the ``explosive'' percolation transition for ``real'' systems have been proposed\cite{Gallos2010,Kumar2011}.
Up to now the systems investigated include networks and lattice bond percolation.

In this paper we extend these investigations to include now site explosive percolation and its finite-size hysteresis properties. We find that site explosive percolation indeed exhibits a very sharp transition with $\beta$ exponent ``effectively'' zero ($\beta/\nu \simeq 0.001$) for the case of sum rule and $\simeq 0.04$ for the case of product rule. We investigate two variants of a reverse ``Achlioptas process (AP1 and AP2)'' and find that the direct and the reverse processes exhibit a hysteresis loop, which surprisingly vanishes in the themodynamic limit.\\ 
The rest of the paper is organized as follows. In Sec. II we describe the site percolation algorithm for the direct and the reverse processes. Moreover, we describe the tools we used to examine the transition and the behavior of the system. In Sec. III we present our results and discuss their significance. Finally, in Sec. IV we sum up our main conclusions.\\

\section{II.Methods}
\begin{figure}[ht]
\centering

\subfigure[]{
   \includegraphics[width=4cm] {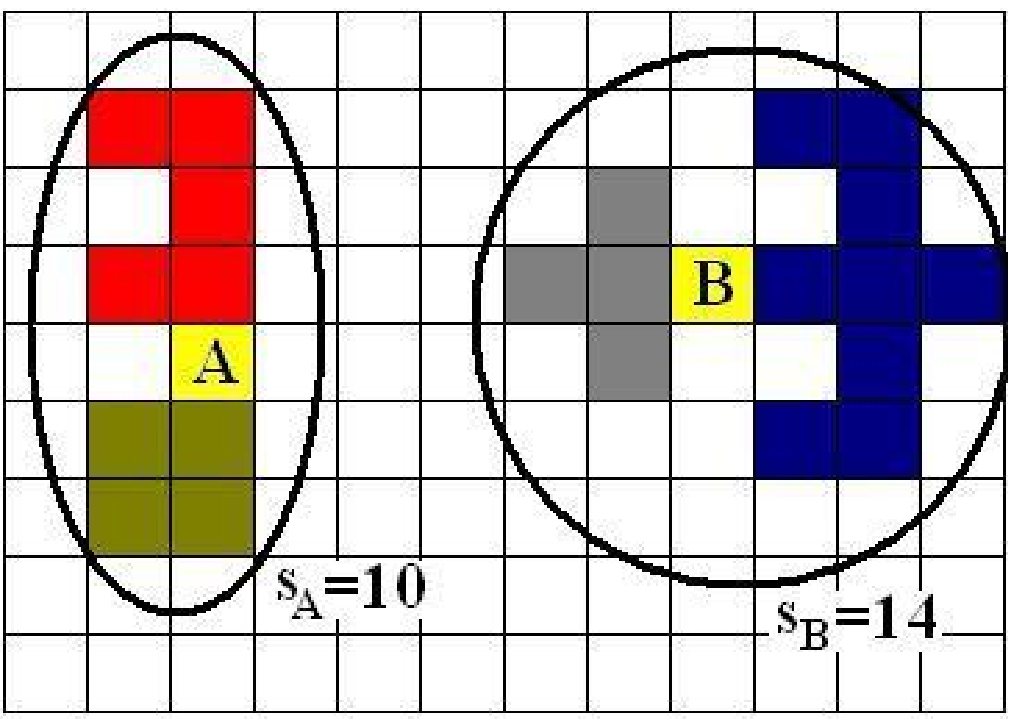}
   \label{Achlioptas1}
 } 
\subfigure[]{
   \includegraphics[width=4cm] {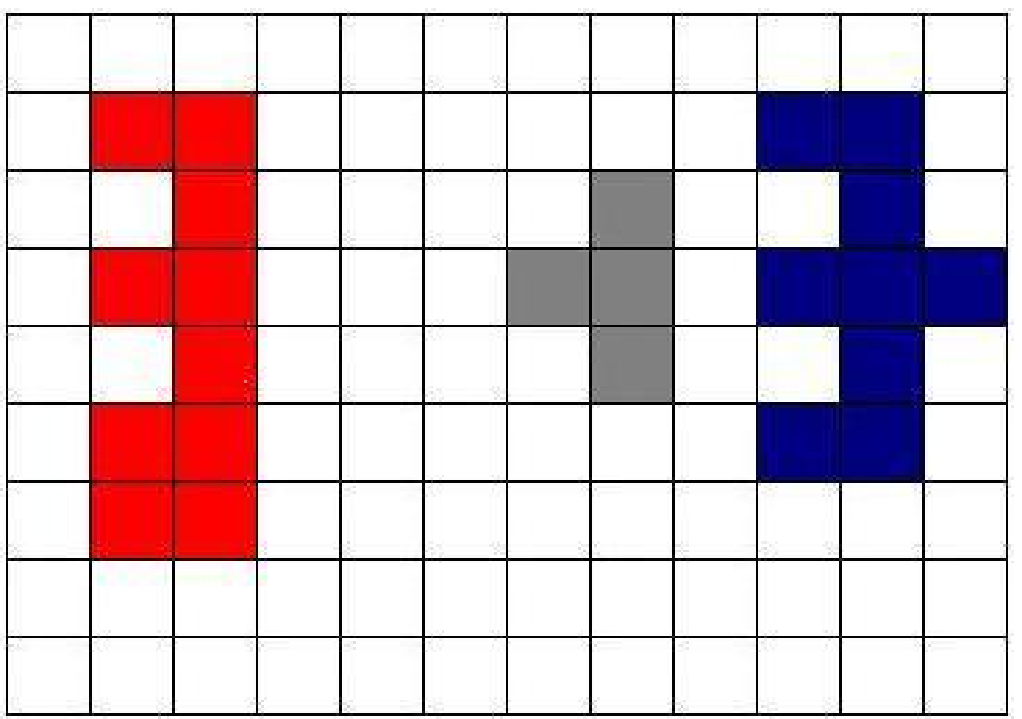}
   \label{Achlioptas2}
 }
 \caption{Achlioptas Process according to the sum rule(APSR) for site percolation . White cells correspond to unoccupied sites while colored cells correspond to occupied sites. Different colors (red,green,gray,blue) indicate different clusters. (a) We randomly select two trial unoccupied sites (yellow), noted by A and B, one at a time. We evaluate the size of the clusters that are formed and contain sites A and B, $s_A$ and $s_B$ respectively. In this example $s_A=10$ and $s_B=14$. (b) According to AP, we keep site A which leads to the smaller cluster and discard site B.}
\label{Achlioptas}
\end{figure}

In the present paper we use Monte Carlo simulations for the site percolation problem on $L \times L$ square lattice with periodic boundary conditions. The algorithm for the case of sum rule (APSR) proceeds as follows:\\
(1) Initially, we start from an empty lattice. We randomly occupy one single site.\\
(2) Next, we randomly select a trial unoccupied site, say A.\\
(3) We calculate the size $s_{A}$ of the resulting cluster in which A belongs to.\\
(4) We remove the trial unoccupied site A and randomly select a trial unoccupied site B, different from A.\\
(5) We calculate the size $s_{B}$ of the resulting cluster in which B belongs to.\\
(6) In case $s_{A} < s_{B}$, site A is permanently occupied and site B is discarded. In case $s_{B} < s_{A}$, site B is permanently occupied and site A is discarded. In case $s_{A} = s_{B}$, we randomly select and permanently occupy either A or B discarding the other. Each time, the number of occupied sites $t$ is incremented by one.\\
(7) We repeat steps (2)-(6) until the entire lattice is covered. For each ``time step'' $t$, we monitor the size of the largest cluster $S_{max}$.\\
In Fig \ref{Achlioptas} we present an illustrative example of the algorithmic process, as has been previously studied. Next, we implemented the product rule (APPR). In this case, we calculate the product of the sizes of the different clusters which will merge after the placement of the new site, A or B. The only changes in the above algorithm are at steps (3) and (5). In order to compare the ``explosive'' percolation transition with the classical site percolation transition, we have simulated random site percolation, using the efficient Newman-Ziff algorithm \cite{Newman_Ziff2}.

\begin{figure}[htbp]
\begin{center}

\subfigure[]{
   \includegraphics[width=3cm] {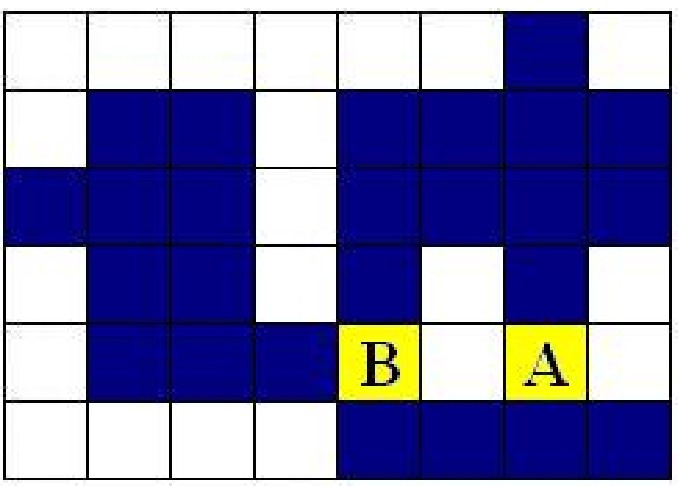}
   \label{reverseAchl1}
 }
 \subfigure[]{
   \includegraphics[width=3cm] {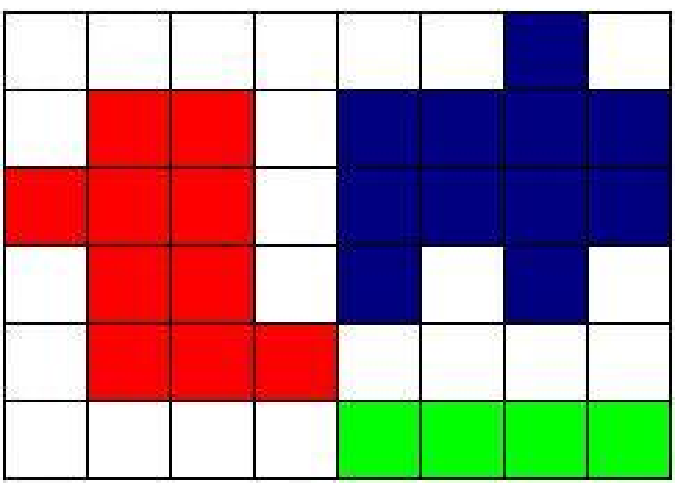}
   \label{reverseAchl2}
 }
\\ 
\subfigure[]{
   \includegraphics[width=3cm] {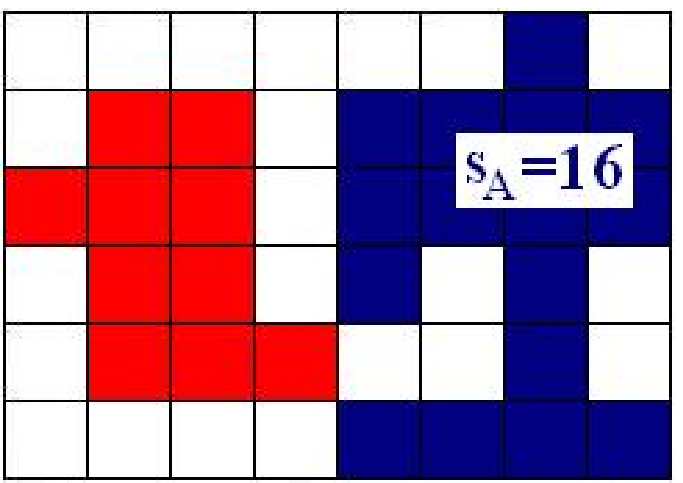}
   \label{reverseAchl3}
 }
\subfigure[]{
   \includegraphics[width=3cm] {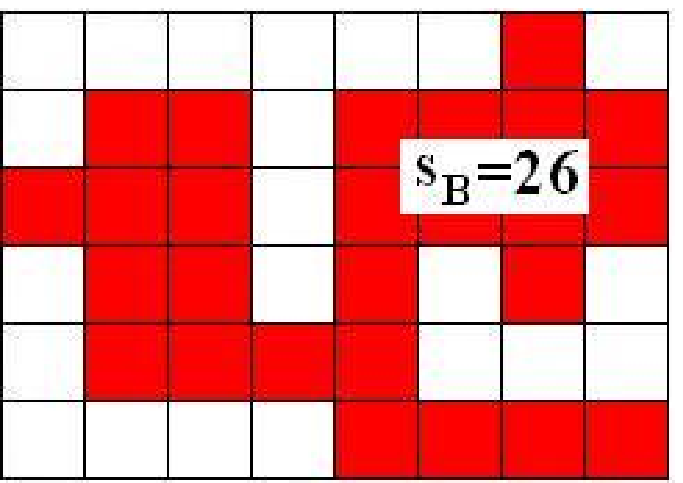}
   \label{reverseAchl4}
 }
 \\
\end{center} 
\caption{Reverse Achlioptas Process (AP1) for site percolation according to the sum rule. Blue is for the occupied sites while white for the unoccupied sites. Initially, the lattice is fully occupied.(a)An instance of the process. We randomly choose two trial sites (yellow), noted as A and B, and remove them from the lattice. (b) The clusters formed after the removal. (c) We place site A again in the lattice and calculate the size of the cluster in which it belongs, $s_A=16$. (d) We do the same as before for the case of site B and calculate $s_B=26$. We remove site A which leads to the formation of the smaller cluster and keep site B.} 
\label{reverseAchlioptas}
\end{figure}

Subsequently, we simulated two variants of reverse ``explosive'' percolation processes, in order to study possible hysteresis phenomena associated with ``explosive percolation''. We have implemented the following algorithm to simulate the reverse process AP1:\\
(1) Initially, we start from a fully occupied lattice.\\
(2) Next, we randomly select two trial occupied sites, say A and B. We remove them temporarily from the lattice.\\
(3) We place again one of the previous sites, say A and calculate the size $s_{A}$ of the resulting cluster in which A belongs to.\\
(4) We go back to step (2) and repeat (3) for the case of site B, calculating the size $s_{B}$ of the resulting cluster in which B belongs to.\\
(5) In case $s_{A} > s_{B}$, site A is reoccupied and site B is permanently discarded. In case $s_{B} > s_{A}$, site B is reoccupied and site A is permanently discarded. In case $s_{A} = s_{B}$, we randomly select and reoccupy either A or B, permanently discarding the other. Each time, the number of occupied sites $t$ is reduced by one.\\
(6) We repeat steps (2)-(5) until the entire lattice is empty. For each ``time step'' $t$, we monitor the size of the largest cluster $S_{max}$.\\
In the above algorithm we did not consider periodic boundary conditions. In Fig \ref{reverseAchlioptas}, we present an illustrative example of the algorithm used. The details are given in the figure caption. We start with a fully occupied lattice. The resulting clusters are shown in Fig \ref{reverseAchl4}.

The reverse process AP2 is a slight modification of the above algorithm for AP1. At step 5 we remove the site which maximizes the resulting cluster. Thus, for AP2 step 5 should be:
(5 -AP2) In case $s_{A} < s_{B}$, site A is reoccupied and site B is permanently discarded. In case $s_{B} < s_{A}$, site B is reoccupied and site A is permanently discarded. In case $s_{A} = s_{B}$, we randomly select and reoccupy either A or B, permanently discarding the other. Each time, the number of occupied sites $t$ is reduced by one.
The rest of the algorithmic steps remain identical to AP1 described above.

Note that the main difference in the selection rule applied in the direct and in the reverse AP1 algorithm is that in the direct case, we add the site which minimizes the resulting cluster, while in the reverse AP1 case, we remove the site which minimizes the resulting cluster. Also, for clarity, we performed the opposite procedure in the reverse case (named reverse AP2 in Fig 7(b)), i.e. we removed the sites which maximized the resulting cluster. 

To track the behavior of the system, we have studied the order parameter $P_{max}$ defined as the ratio of the sites that belong to the largest cluster $S_{max}$ to the total number of sites in the lattice. For continuous transitions, $P_{max}$ follows the scaling relation \cite{Bunde_BOOK,Stauffer}:
\begin{equation}
P_{max}=L^{-\beta/\nu}F[(p-p_{c})L^{1/\nu}]
\label{eq1}
\end{equation}in the vicinity of $p_c$. 
For the present analysis, we assume that the order parameter of the transition follows the conventional well-known finite size scaling. Recently, it has been suggested \cite{Grassberger2011} that ``explosive'' transitions are continuous, but with an unconventional finite size scaling. We do not examine this possibility in the present manuscript. 
Moreover, we monitor the standard deviation of the size of the largest cluster $S_{max}$, defined as
\begin{equation}
\chi=\sqrt{<S_{max}^{2}>-<S_{max}>^{2}}
\label{eq2}
\end{equation} which has recently been proposed \cite{ziff2010explosive,araujoHermann2010explosive} as a criterion to discern between continuous and ``explosive'' transitions.

We also employed the $\Delta$ criterion proposed in \cite{achlioptasDSouza2009} which quantifies the transition width of a system with size $N$. It is defined as $\Delta= t_{1}- t_{0}$,  where $t_{1}$ is the lowest value for which $S_{max}>0.5 N$ and $t_{0}$ the lowest value for which $S_{max}> \sqrt{N}$. Achlioptas et al. stated that for continuous (i.e. random) percolation, this quantity should scale as $\Delta/N \rightarrow 0$ for AP and $\Delta/N \rightarrow const$ for RP as $N \rightarrow \infty$. However, this simple scaling relation does not seem to hold for bond percolation on square lattices \cite{ziff2009explosive}.

The data given in the following figures are averages over 1000 realizations.

\section{III.RESULTS AND DISCUSSION}

\begin{figure}[ht]
\centering

\subfigure[]{
   \includegraphics[width=7cm] {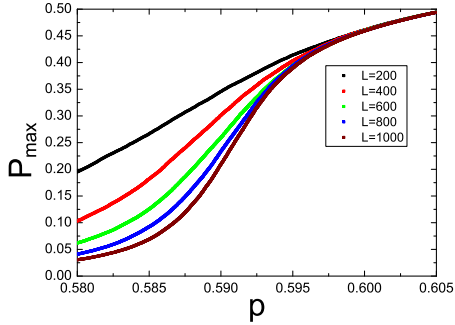}
   \label{classicalPmax}
 } 
 \\
\subfigure[]{
   \includegraphics[width=7cm] {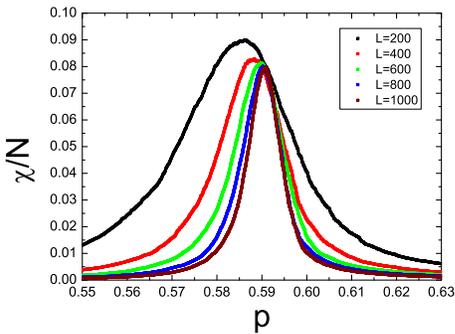}
   \label{classicalx}
 } 
 
\caption{ Random Percolation (RP).(a) Plot of $P_{max}$ as a function of $p$ for different linear sizes $L$. (b) Plot of $\chi/N$ as a function of $p$ for the same sizes $L$ as in (a). It is evident that the peaks of the curves decay as $L$ increases. Colors are: $L=200$(black), $L=400$(red), $L=600$(green), $L=800$(blue), $L=1000$(wine)}
\label{Classical}
\end{figure}

\begin{figure}[ht]
\centering

\subfigure[]{
   \includegraphics[width=7cm] {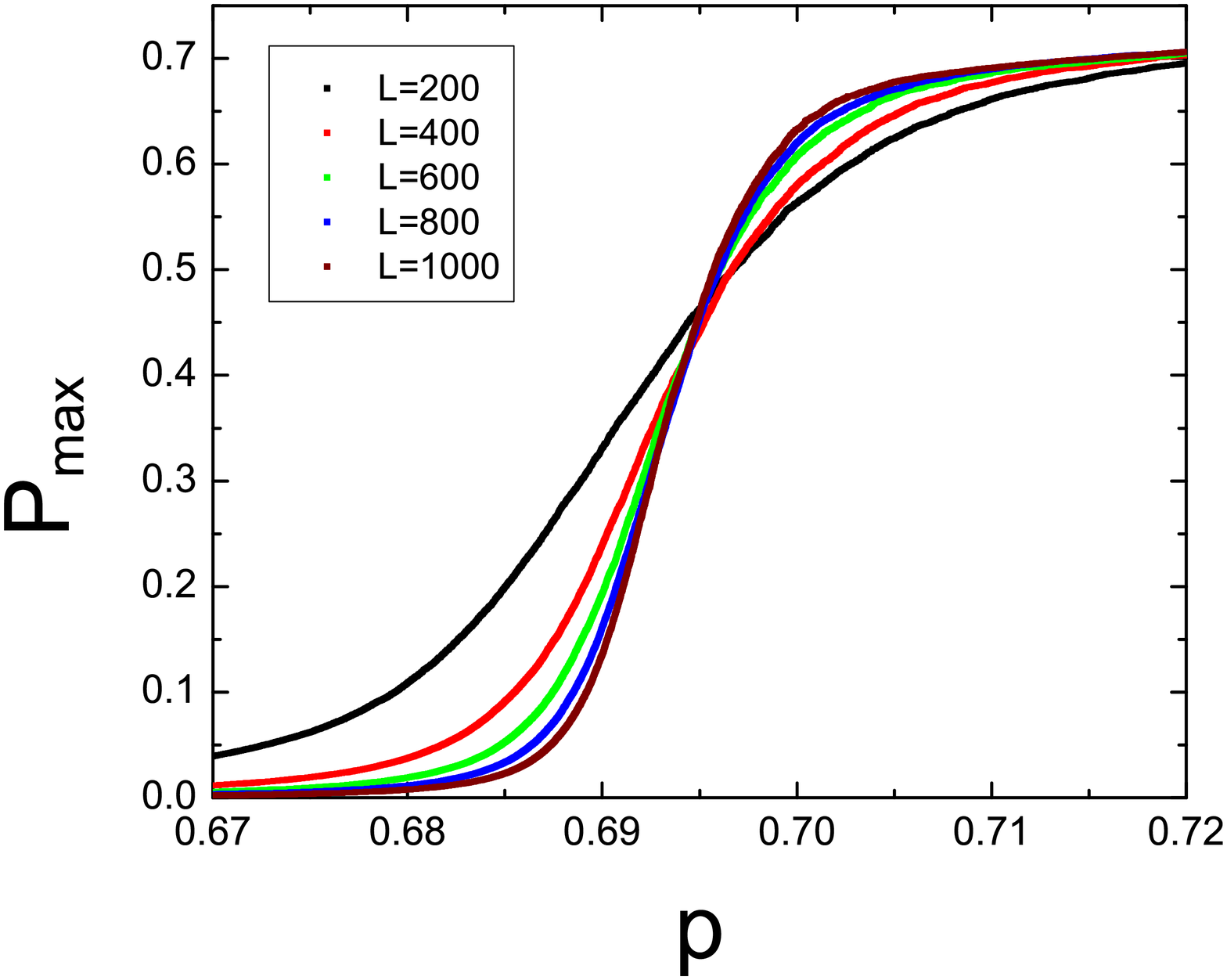}
   \label{achlioptasPmax}
 } 
 \\
\subfigure[]{
   \includegraphics[width=7cm] {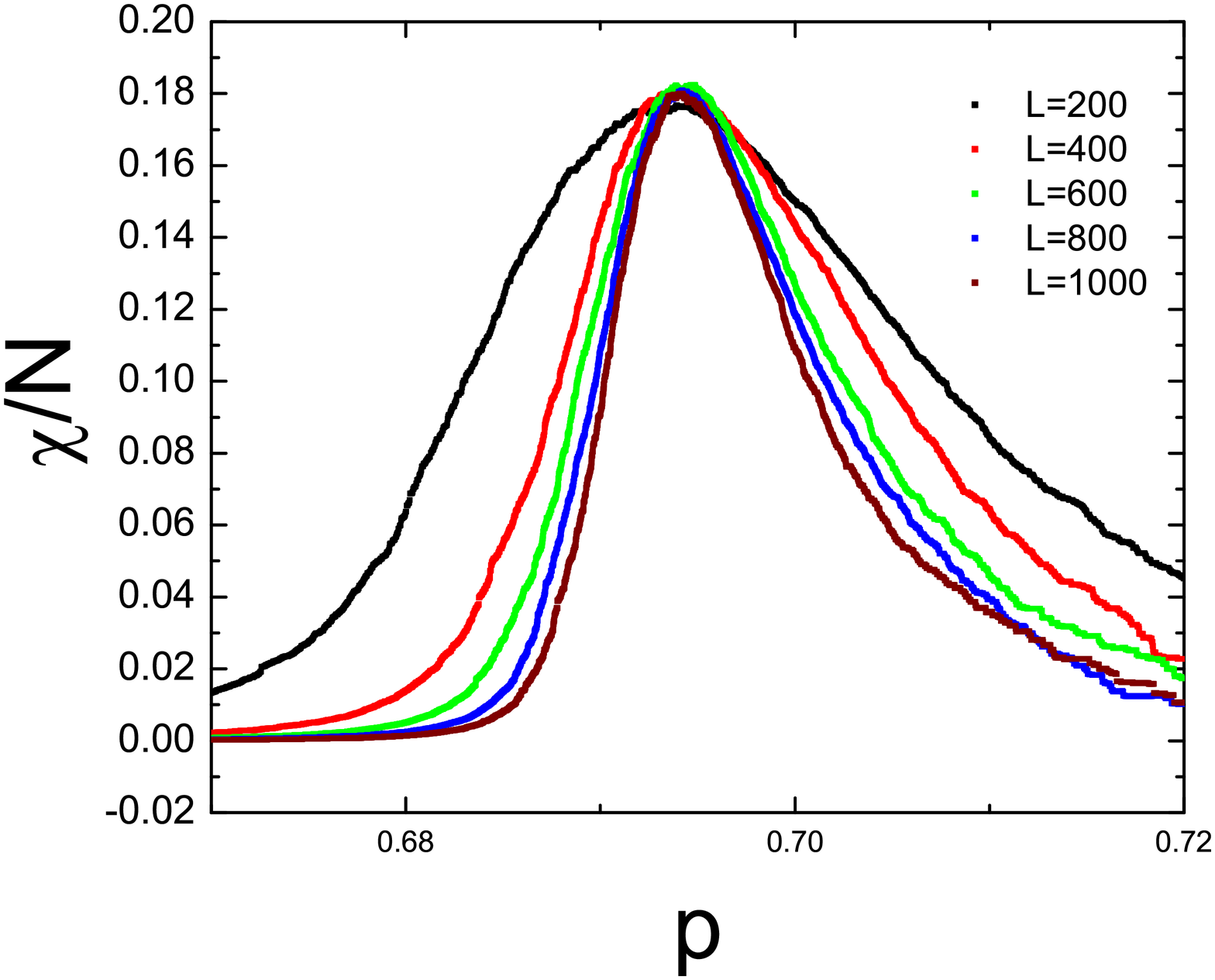}
   \label{achlioptasx}
 } 
 
\caption{Achlioptas Process with sum rule (APSR). (a)Plot of $P_{max}$ as a function of $p$ for different linear sizes $L$ for site percolation. We observe that all plotted lines pass through a single point at $p \simeq 0.695$. Moreover, we see a more abrupt jump of $P_{max}$ compared to \ref{classicalPmax}. (b)Plot of $\chi/N$ as a function of $p$ for the same sizes $L$ as in (a). Peaks are almost constant. Colors are: $L=200$(black), $L=400$(red), $L=600$(green), $L=800$(blue), $L=1000$(wine)}
\label{AchlioptasSR}
\end{figure}

\begin{figure}[ht]
\centering

\subfigure[]{
   \includegraphics[width=7cm] {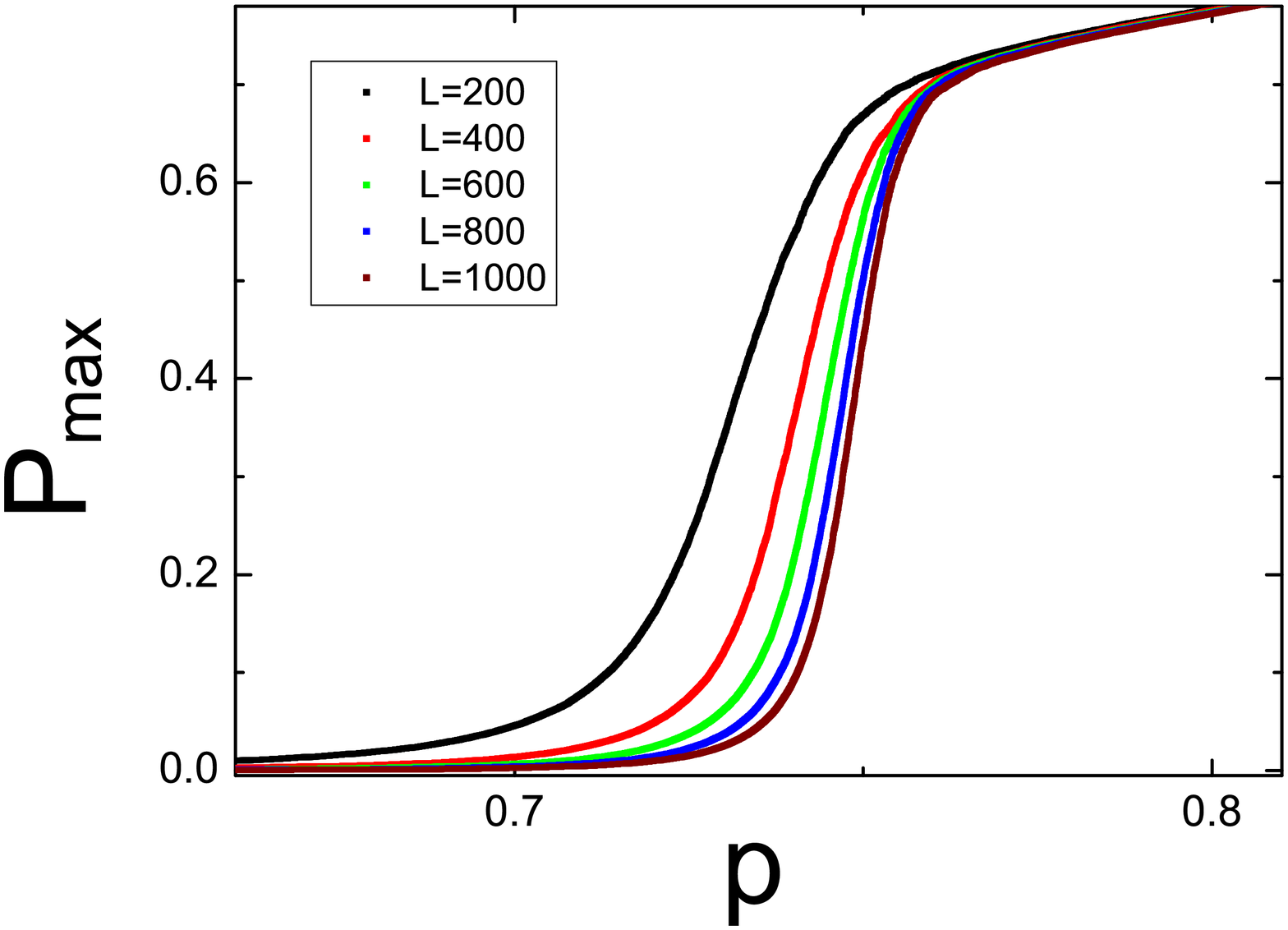}
   \label{achlioptasPmaxPR}
 } 
\subfigure[]{
   \includegraphics[width=7cm] {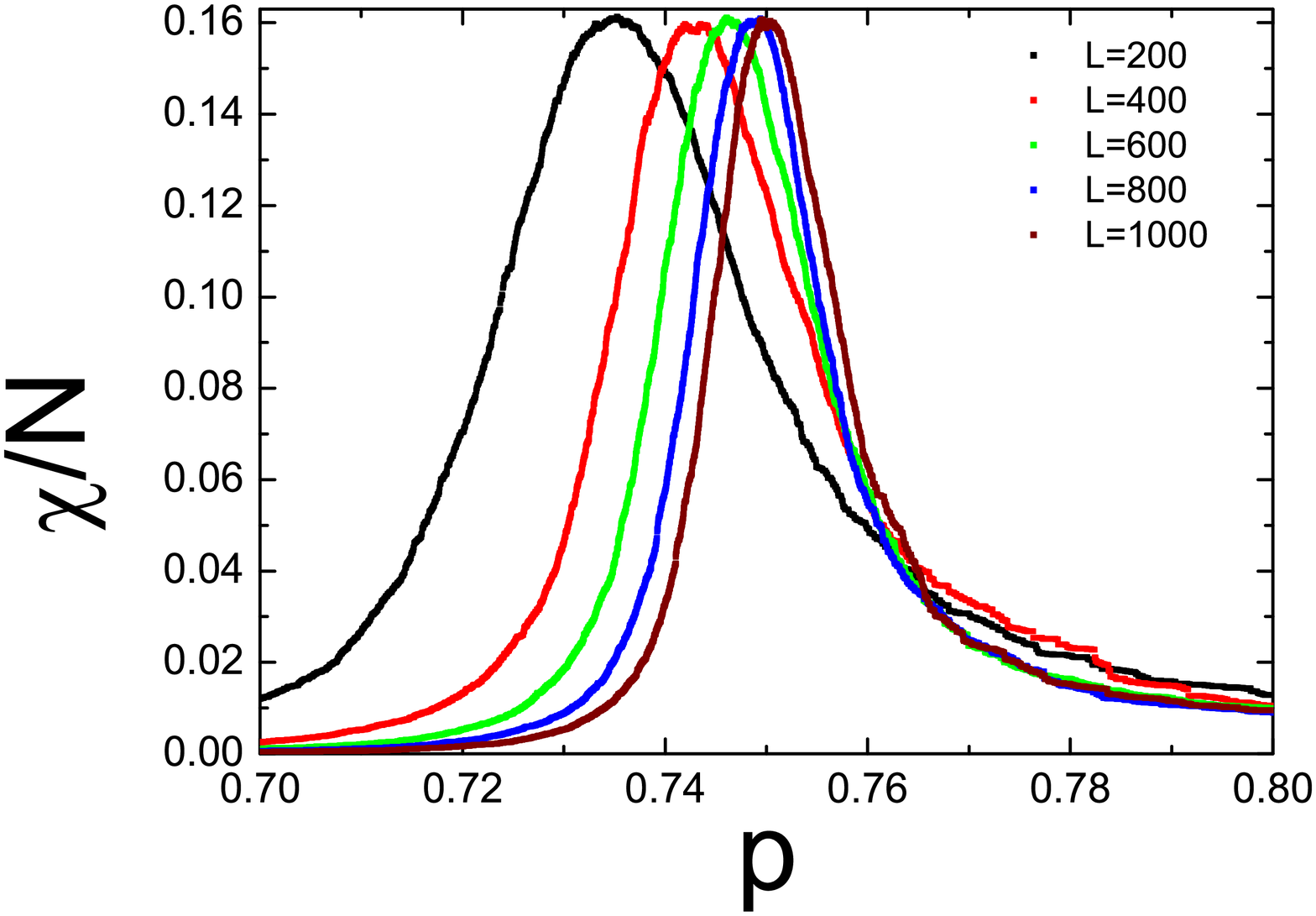}
   \label{achlioptasxPR}
 }

\caption{Achlioptas Process with product rule (APPR). (a) Plot of $P_{max}$ as a function of $p$ for different linear sizes $L$ for site percolation.(b) Plot of $\chi/N$ as a function of $p$ for the same sizes $L$ as in (a). Colors are: $L=200$(black), $L=400$(red), $L=600$(green), $L=800$(blue), $L=1000$(wine)}
\label{achlioptasPR}
\end{figure}

In this section we present the results obtained from the calculations. In Figs \ref{classicalPmax}, \ref{achlioptasPmax} and \ref{achlioptasPmaxPR} we plot the order parameter $P_{max}$ as a function of $p$, for the case of RP, APSR and APPR models, respectively, for five different system sizes, namely $L=200 (black), 400(red), 600(green), 800(blue)$ and $1000(wine)$. For the case of RP model, we have recovered the correct critical exponents ($\beta/\nu \simeq 0.106$ and $1/\nu \simeq 0.75$). A comparison between Figs \ref{classicalPmax},\ref{achlioptasPmax} and \ref{achlioptasPmaxPR} shows that $P_{max}$ changes more abruptly in the case of AP models, as expected from an ``explosive'' transition. Moreover, in the APSR case, we observe that at $p \simeq 0.695$ all lines for the plotted system sizes seem to be crossing roughly at the same point. This $p$ value coincides with our estimations of the transition point $p_c$ for the APSR model, using the method of the second largest cluster as well as the value $p$ at which the $\chi$ per lattice site ($\chi/N$) reaches its maximum value. Using standard finite size scaling techniques, we find  that $\beta/\nu \simeq 0.001$. More precisely, Eq. \ref{eq1} implies that $L^{\beta/\nu}P_{max}=F[(p-p_c)L^{1/\nu}]$ and, thus, curves of $L^{\beta/\nu}P_{max}$ versus p for different system sizes should cross at a single point at $p=p_c$. We have used our Monte Carlo data for $P_{max}$ vs $p$ for 5 different sizes $L$, to create 5 functions $g_i = L^{\beta/\nu}P_{max}$ and used minimization software to determine the values $\beta/\nu$ and $p$ which minimize the sum of the squares of the pairwise differences $\Lambda \equiv \Lambda(\beta/\nu,p)=\sum_{i\neq j}{{(g_i-g_j)}^{2}}$. 

\begin{figure}[ht]
\centering

\subfigure[]{
   \includegraphics[width=8cm] {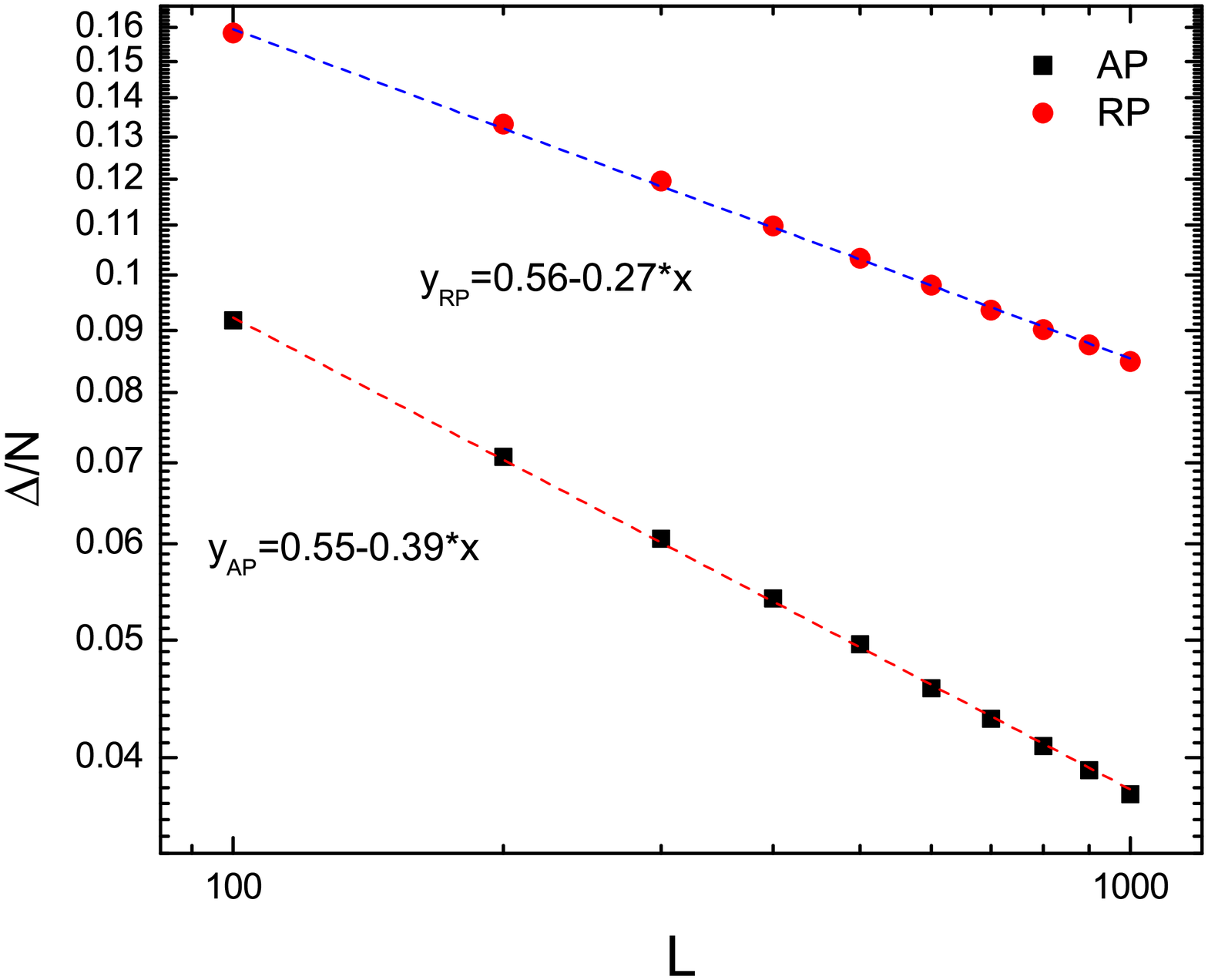}
   \label{DELTA}
 } 
\subfigure[]{
   \includegraphics[width=8cm] {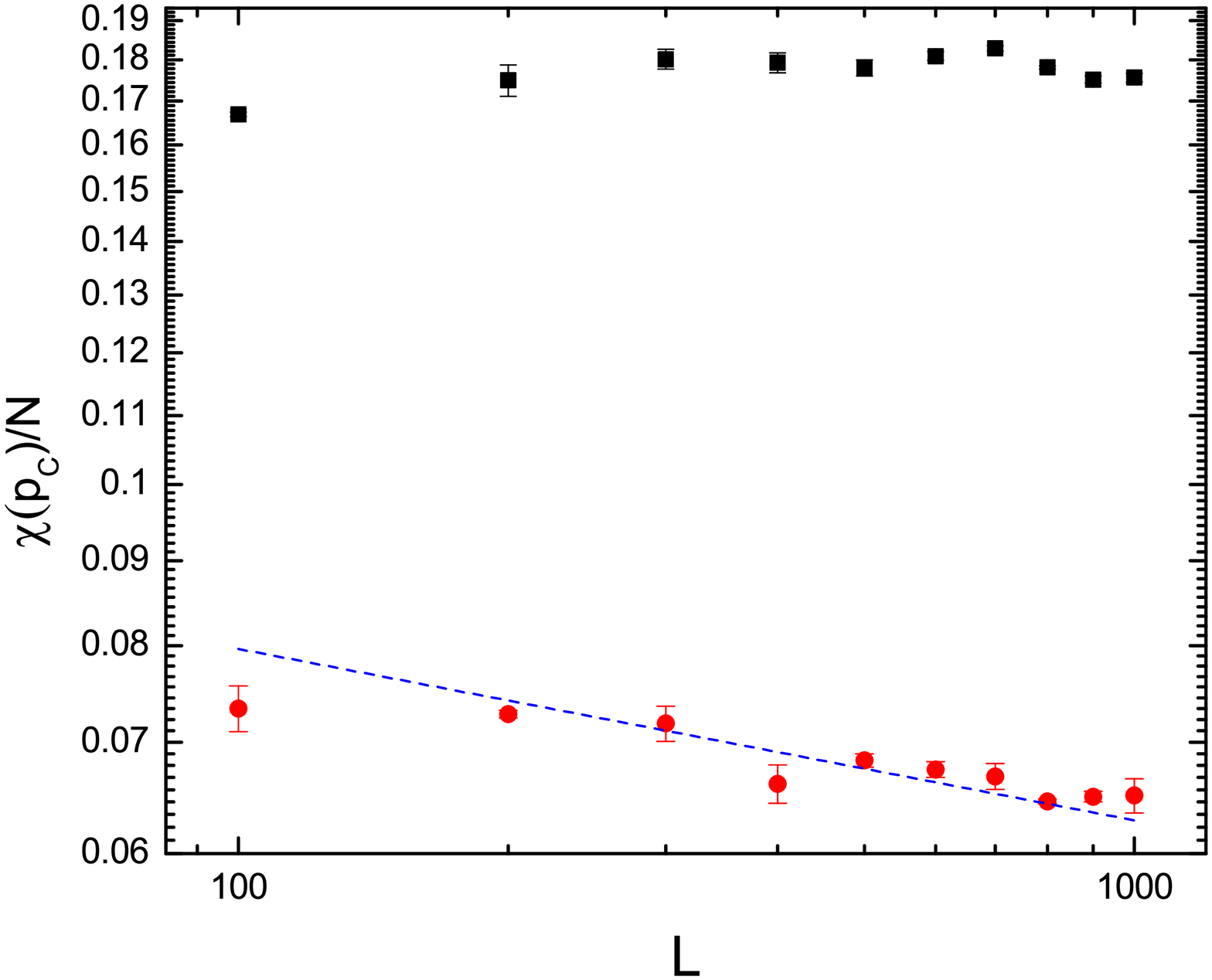}
   \label{xpc}
 }
 
\caption{(a) $\Delta/N$ as a function of $L$ for RP (red dots) and APSR (black squares). Both scale as a power-law, with $L^{-0.27}$ for RP case and $L^{-0.39}$ for APSR case. (b) Plot of $\chi(p_{c})/N$ as a function of $L$ for RP (red dots) and APSR (black squares). Blue dashed line is for the theoretical prediction of the variation of $\chi(p_{c})/N$ for the case of RP. We observe a power-law decay for the RP case. In the APSR case, $\chi(p_c)/N$ initially increases and reaches a plateau for large systems. Both plots are in logarithmic scale.} 
\label{xpcDELTA}
\end{figure}

Ideally, $\Lambda(\beta/\nu,p_c)$ should be exactly zero, but since numerical errors cannot be avoided, we consider that the values of $\beta/\nu$ and $p$ that minimize $\Lambda$ provide an accurate estimation of the actual values of $\beta/\nu$ and $p_c$. Such a small value for the $\beta/\nu$ as the one observed for the APSR, is very unusual for a continuous transition. Of course, it cannot be determined whether a transition is actually discontinuous or continuous with an extremely small $\beta$, as already discussed in ref \cite{Dorogovtsev2010}. It is, however, very interesting that our calculated value $\beta/\nu$ is different from the values calculated by Ziff \cite{ziff2010explosive} and Radicchi \textit{et al.} \cite{Radicchi2010explosive} for bond percolation. They also could not reach a sharp conclusion regarding the nature of the transition. In contrast, in  \cite{Dorogovtsev2010,Grassberger2011} numerical evidence is provided for the continuity of ``explosive'' percolation transition. Our findings indicate that in case the ``explosive'' transition is continuous, the site ``explosive'' percolation and bond ``explosive'' percolation belong to different universality classes. This result is both intriguing and unexpected, since in the classical case, it is well-known that the critical exponents do not depend on the structure of the lattice (e.g., square or triangular) or on the type of percolation (site, bond or even continuum)\cite{Stauffer}. We must emphasize however, that Grassberger \textit{et al.} \cite{Grassberger2011} suggest that ``explosive'' percolation follows an unconventional finite size scaling.\\
In Figs \ref{classicalx} \ref{achlioptasx} and \ref{achlioptasxPR}, we plot the normalized standard deviation $\chi/N$ of the size of the largest cluster $S_{max}$, for the same five system sizes $L$ as above, for the RP, APSR and APPR case, respectively. In all cases, $\chi/N$ exhibits a sharp maximum at $p_c$. However, in the RP case, we observe that the maximum value is a decreasing function of $L$, while in the cases of APSR and APPR, it is nearly constant for large systems (see also Fig \ref{xpc} below for APSR).

As discussed above, Fig \ref{achlioptasPR} shows our results from the product rule simulations (APPR). Using the minimization technique previously described, we calculated the value of $p_c$ to be $p_c=0.756\pm0.006$ in accordance to \cite{KimYup2011} within error bounds. As for $\beta/\nu$ we calculated it to be $\beta/\nu = 0.04 \pm0.02$.

In Fig \ref{DELTA}, we plot the quantity $\Delta/N$ as a function of the system size $L$, both for RP (red dots) and APSR (black squares) models. We observe a power-law scaling for both cases, with an exponent $\omega_{RP}=-0.27$ and $\omega_{AP}=-0.39$. As expected, and in agreement with previous results reported by Achlioptas \textit{et al.} for the case of random networks \cite{achlioptasDSouza2009} and by Ziff for bond percolation on lattices \cite{ziff2009explosive}, we find that the exponent for the AP case is much smaller than the exponent for the RP case, indicating a sharper transition for the former. Again, we observe that the exponent $\omega_{AP}$ for the site ``explosive'' percolation is different from the exponent for the case of bond ``explosive'' percolation reported by Ziff in \cite{ziff2009explosive}, providing an additional indication that site and bond ``explosive'' percolation belong to different universality classes.

Next, we study how the quantity $\chi/N$ scales with the system size $L$ for the classical and the ``explosive'' case. For the RP case, we expect, by a simple scaling argument, that the quantity $\chi/N$ decays as $L^{\gamma/{2\nu}-1}$. It is known that the variance of the order parameter $P_{max}$ is related to the susceptibility exponent $\gamma$ and scales at $p_c$ as $var(P_{max})\thicksim L^{\gamma/\nu}$. Thus, $var(S_{max}) \thicksim L^{{\gamma/\nu}+ 2}$. The standard deviation of the quantity $S_{max}$, $\chi$, is the square root of the variance and, thus, $\chi \thicksim L^{{\gamma/2\nu} + 1}$. And, finally, $\chi/N \thicksim L^{\gamma/{2\nu}-1}$. For the RP process, the values of $\gamma$ and $\nu$ are known exactly and they are $\gamma = 43/18$ and $\nu = 4/3$. This leads to the conclusion that $\chi/N \thicksim -5/48 \simeq -0.1$. In Fig \ref{xpc}, we plot $\chi(p_c)/N$ as a function of $L$ for RP (red dots) and for APSR (black squares). Points are Monte Carlo simulation results, while the dotted line is the theoretically expected slope for the RP case. We did observe that the RP case data show a power-law decay, while for the APSR case, we observe an initial increase of $\chi/N$ followed by a plateau for large system sizes.

\begin{figure}[ht]
\centering

\subfigure[]{
   \includegraphics[width=8cm] {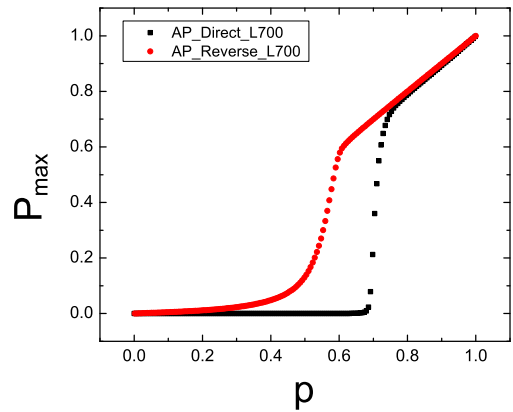}
   \label{hysteresis1}
 } 
\subfigure[]{
   \includegraphics[width=9cm] {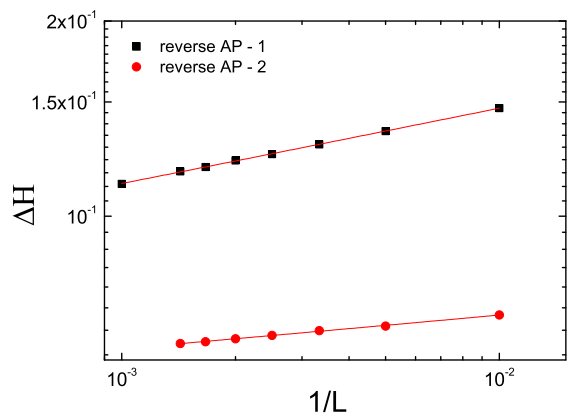}
   \label{hysteresis2}
 }

\caption{(a) Plot of the direct (black squares) and reverse (red dots)AP1 for a $700$ x $700$ lattice. The hysteresis loop is evident. (b) The area of the hysteresis loop $\Delta H$ as a function of $1/L$ (logarithmic scale). Black squares are for reverse AP1 and red dots are for reverse AP2, as described in Sec. II. For reverse AP1, $\Delta H \thicksim (1/L)^{0.125}$ and for reverse AP2 $\Delta H \thicksim (1/L)^{0.069}$. It is clear that the reverse and forward processes become equivalent in the thermodynamic limit ($L \rightarrow \infty$).}
\label{hysteresis}
\end{figure}

Finally, in Fig \ref{hysteresis}, we present the results for the direct and the reverse AP processes (AP1 and AP2) described in Sec. II. In Fig \ref{hysteresis1}, we plot $P_{max}$ as a function of $p$ for a $700 \times 700$ lattice, both for the direct and the reverse AP1 case. We observe that for a finite system, there exists a hysteresis loop. Of course, no such loop exists in the classical RP case, where the process is fully reversible. We are interested in determining the scaling properties of this loop. We should emphasize that in contrast to thermal processes, we cannot associate the observed hysteresis loops with metastability and first order transitions. We merely use the ``hysteresis'' concept as a means to investigate and compare different but related processes. In Fig \ref{hysteresis2}, we plot the area $\Delta H$ surrounded by the curves corresponding to the the direct and reverse processes for various sizes $L$ as a function of $1/L$. Black squares are for the reverse AP1 and red dots are for the reverse AP2, as described in Sec. II. The fitting suggests that $\Delta H \thicksim (1/L)^{0.125}$ for the former and $\Delta H \thicksim (1/L)^{0.069}$ for the latter. Thus, we find that, although in finite systems, the reverse processes are different from the direct process, they become equivalent in the thermodynamic limit ($L \rightarrow \infty$).  

\section{IV.Conclusions}
We have studied random (RP) and Achlioptas processes (APSR and APPR) for site percolation on 2D square lattices. We report the critical point for the APSR as $p_c=0.695$ and $p_c=0.756$ for APPR. We have shown that the order parameter $P_{max}$ changes more abruptly for the case of AP processes compared to the RP case. Using standard finite size scaling techniques, we find indications that ``explosive'' site and bond percolation may belong to different universality classes. A qualitatively similar behavior was also observed for $\Delta/N$, where the exponents for ``explosive'' site and bond percolation are different. Finally, the quantity $\chi/N$ shows at $p_c$ a power-law decay for the case of RP, while it appears to be nearly constant for large systems in the case of AP models (APSR and APPR). 

We also have studied the direct and two variants of a reverse AP process. We have shown that, although in finite size systems, the reverse processes are different from the direct process, they become both equivalent in the thermodynamic limit.

\begin{acknowledgments}
\textbf{Acknowledgments}\\
N.B. acknowledges financial support of Public Benefit Foundation Alexander S. Onassis.\\
\end{acknowledgments}

\end{document}